\documentclass[10pt,twocolumn,a4paper,
               aps,prl,
               superscriptaddress,
               showpacs,
               amsmath,
               amssymb,
              ]{revtex4-1}

\usepackage{graphicx}
\usepackage{xspace}
\usepackage{xcolor}
\usepackage[normalem]{ulem}

\begin{document}

\date{\today}

\title{Mode Coupling of Phonons in a Dense One-Dimensional Microfluidic Crystal}

\author{Jean-Baptiste Fleury}
\thanks{J.-B.~F. and U.~D.~S. contributed equally to this work.}
\affiliation{Experimental Physics, Saarland University, 66123 Saarbr\"ucken, Germany}

\author{Ulf D. Schiller}
\thanks{J.-B.~F. and U.~D.~S. contributed equally to this work.}
\affiliation{Theoretical Soft Matter and Biophysics, Institute of Complex Systems, Forschungszentrum J\"ulich, 52425 J\"ulich, Germany}

\author{Shashi Thutupalli}
\affiliation{Max Planck Institute for Dynamics and Self-Organization, 37077 G\"ottingen, Germany}

\author{Gerhard Gompper}
\affiliation{Theoretical Soft Matter and Biophysics, Institute of Complex Systems, Forschungszentrum J\"ulich, 52425 J\"ulich, Germany}

\author{Ralf Seemann}
\affiliation{Experimental Physics, Saarland University, 66123 Saarbr\"ucken, Germany}
\affiliation{Max Planck Institute for Dynamics and Self-Organization, 37077 G\"ottingen, Germany}


\begin{abstract}
  Long-living coupled transverse and longitudinal phonon modes are
  explored in dense and regular arrangements of flat microfluidic
  droplets. The collective oscillations are driven by hydrodynamic
  interactions between the confined droplets and can be excited in a
  controlled way. Experimental results are quantitatively compared to
  simulation results obtained by multi-particle collision dynamics.
  The observed transverse modes are acoustic phonons and can be
  described by a linearized far-field theory, whereas the longitudinal
  modes arise from a non-linear mode coupling due to the lateral
  variation of the flow field under confinement.
\end{abstract}

\maketitle

\emph{Introduction.}
Hydrodynamic interactions of small objects in driven non-equilibrium
systems lead to complex phenomena such as strong correlations in
sedimenting suspensions~\cite{Ramaswamy2001}, shock-waves in
two-dimensional disordered ensembles of droplets~\cite{Beatus2009},
or cluster formation and alignment of red blood cells in micro-capillaries
\cite{McWhirter2009}.
Non-equilibrium many-body phenomena can be favorably studied in
microfluidic systems using ordered arrangements of flowing droplets
(microfluidic
crystals)~\cite{Garstecki2006,Baron2008,Dreyfus2003,Desreumaux2012}.
In pioneering studies, Beatus et
al.~\cite{Beatus2006,Beatus2007,Beatus2012} observed collective
vibrations in dilute droplet trains flowing in a microfluidic
Hele-Shaw configuration. These oscillations have been characterized as
acoustic phonons with properties similar to lattice modes in
dusty-plasma crystals~\cite{Vladimirov1997,Ivlev2000}.
%\dcom{[US] I suggest to use the two more general references here, and
%\cite{Couedel2010,Liu2010} in the text below.}
Their dispersion relation has been obtained to linear-order in terms
of a small-amplitude expansion of the ensemble
flow~\cite{Beatus2006,Beatus2007}. However, in contrast to solid-state
phonons that can be excited in a controlled way by mechanical or
optical means, microfluidic phonons have so far been observed only
when excited by thermal noise or defects in
microchannels~\cite{Beatus2006,Beatus2007,Beatus2012}.  Possible
non-linear effects and interactions between phonon modes could not be
studied quantitatively in these settings \cite{Beatus2012,Liu2012}.

In this letter, we broaden the possibility to study hydrodynamically
mediated collective phenomena by exploring dense droplet systems. In
these systems, long-living transverse phonon modes and coupled
longitudinal oscillations are experimentally excited and explored in
an one-dimensional flowing microfluidic crystal. The experimental
measurements are compared to existing analytic results based on a
linearized theory, and to results from computer simulations using a
mesoscale hydrodynamics approach which is able to reproduce the full
non-linear interactions. The excited transverse phonon modes agree
with the linearized far-field approximation
\cite{Beatus2007,Beatus2012}. However, the observed longitudinal modes
reveal a non-linear coupling to the transverse modes with a dispersion
relation that is beyond the existing analytic theory.

\begin{figure}
 \includegraphics[width=.95\columnwidth]{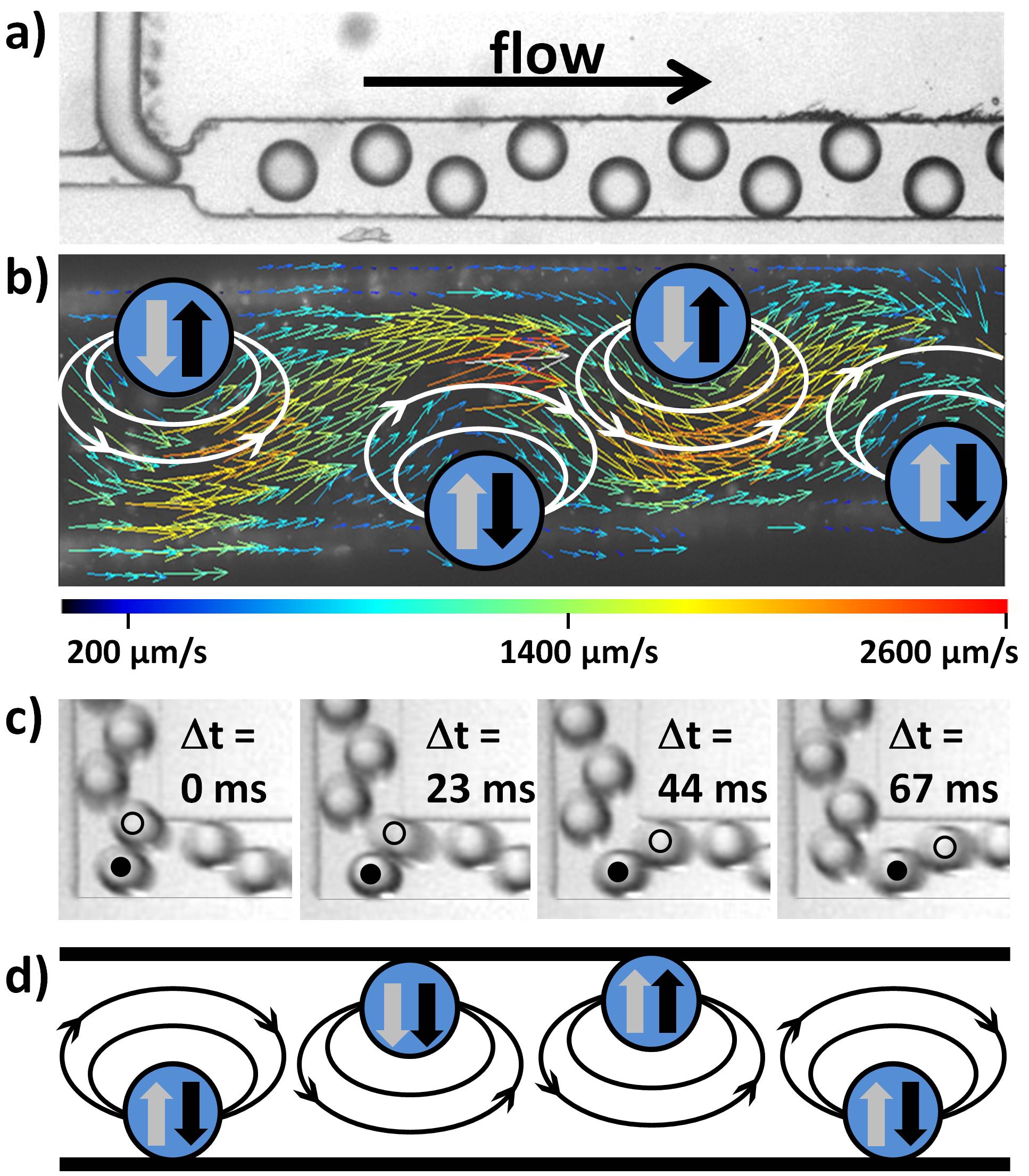}
 \caption{a) Generation of microfluidic crystals using a step
    junction.
    b) Flow field determined by particle image velocimetry in the lab
    frame \cite{supplement}.  The dipolar flow fields around each
    droplet in the co-moving frame of the droplets are sketched by
    white arrows. The transverse forces resulting from leading and
    trailing droplets are shown as gray and black arrows, respectively.
    c) Microscopy time series showing the droplet reorganization at a
    $90^\circ$ bend. The droplet marked with a dot is repelled from the
    corner due to its hydrodynamic image~\cite{Beatus2006,Beatus2009,Beatus2012}
    and the trailing droplet marked with a circle is squeezed between the
    leading droplet and the wall and propelled longitudinally out of the
    corner.
    d) Sketch indicating the dipolar flow fields and the resulting
    transverse forces in a zigzag arrangement with one defect.}
  \label{fig:Picture1}
\end{figure}

\emph{Experimental Setup and Simulation Method.}
Microfluidic devices were fabricated using standard soft lithographic
protocols \cite{Duffy1998,Chokkalingam2008} and the flow rates were
volume-controlled using syringe pumps. Mono-disperse water droplets
are generated in n-hexadecane ($\rho=773\,\text{kg/m}^3$, $\eta =
3\,\text{mPa\,s}$) with 2\,wt\% of the surfactant Span\,80 using a
step geometry \cite{Priest2006,Chokkalingam2008},
cf. Fig.~\ref{fig:Picture1}a. The microchannel has uniform height and
width of $H \times W \approx 120 \times 210\,\mu\text{m}^2$. Typical
flow velocities are $u_{d} \approx 250\,\mu\text{m/s}$ for the
droplet, and $u_{oil} \approx 500\,\mu\text{m/s}$ for the continuous
oil phase.  The corresponding Reynolds and Peclet number are $Re =
\rho u_\text{oil} R/\eta \approx 10^{-2}$ and $Pe = u_\text{oil} R/D
\approx 10^{8}$, where $R$ is the droplet radius and
$D\approx10^{-12}\text{cm}^2\text{/s}$ the diffusion coefficient of a
droplet calculated with the formulas given in~\cite{Beatus2012}.

Computer simulations were conducted using multi-particle collision
dynamics
(MPC)~\cite{Malevanets1999,Goetze2007,Kapral2008,Gompper2009}. In this
mesoscopic method, hydrodynamics are fully reproduced through the
dynamics of point-like solvent particles \cite{supplement}. The
flattened droplets are modeled as discs in two-dimensional channel
flow \mbox{\cite{Goetze2011}}. The collective oscillations are modeled
by an initial zigzag arrangement of droplets with a given periodicity.
Droplet and flow velocities were tuned to give a Reynolds number $Re
\approx 10^{-3}$ in the range of the experiments. Since MPC includes
thermal fluctuations, the Peclet number $Pe \approx 10^2$ for feasible
droplet and flow velocities is significantly smaller than in the
experiments. Consequently, a broader range of frequencies is excited
in the simulations which helps to reveal the full dispersion relation
of the phonon modes as will be discussed below.

\emph{Controlled Excitation of Collective Oscillations.}
The size of the generated droplets exceeds the height of the
microfluidic channel and the droplets are flattened. Due to friction
with the confining top and bottom wall, the flat droplets move slower
than the surrounding liquid phase leading to a two-dimensional dipolar
flow field around each droplet, cf. Fig.~\ref{fig:Picture1}b. Initially,
the produced droplets form a stable zigzag configuration where the
forces from the dipolar flow fields from the neighboring droplets cancel
out. When this droplet arrangement is guided around a bend, a defect in
the ordering can be achieved, as displayed in Fig.~\ref{fig:Picture1}c.
In the resulting droplet arrangement the zigzag symmetry is broken and
some droplets experience a net hydrodynamic force and are moved towards
the opposite channel wall, see Fig.~\ref{fig:Picture1}d. Following this
transverse movement a single defect propagates backwards in the
co-moving frame of the droplets, cf. \cite{supplement}.

\begin{figure}
  \includegraphics[width=.95\columnwidth]{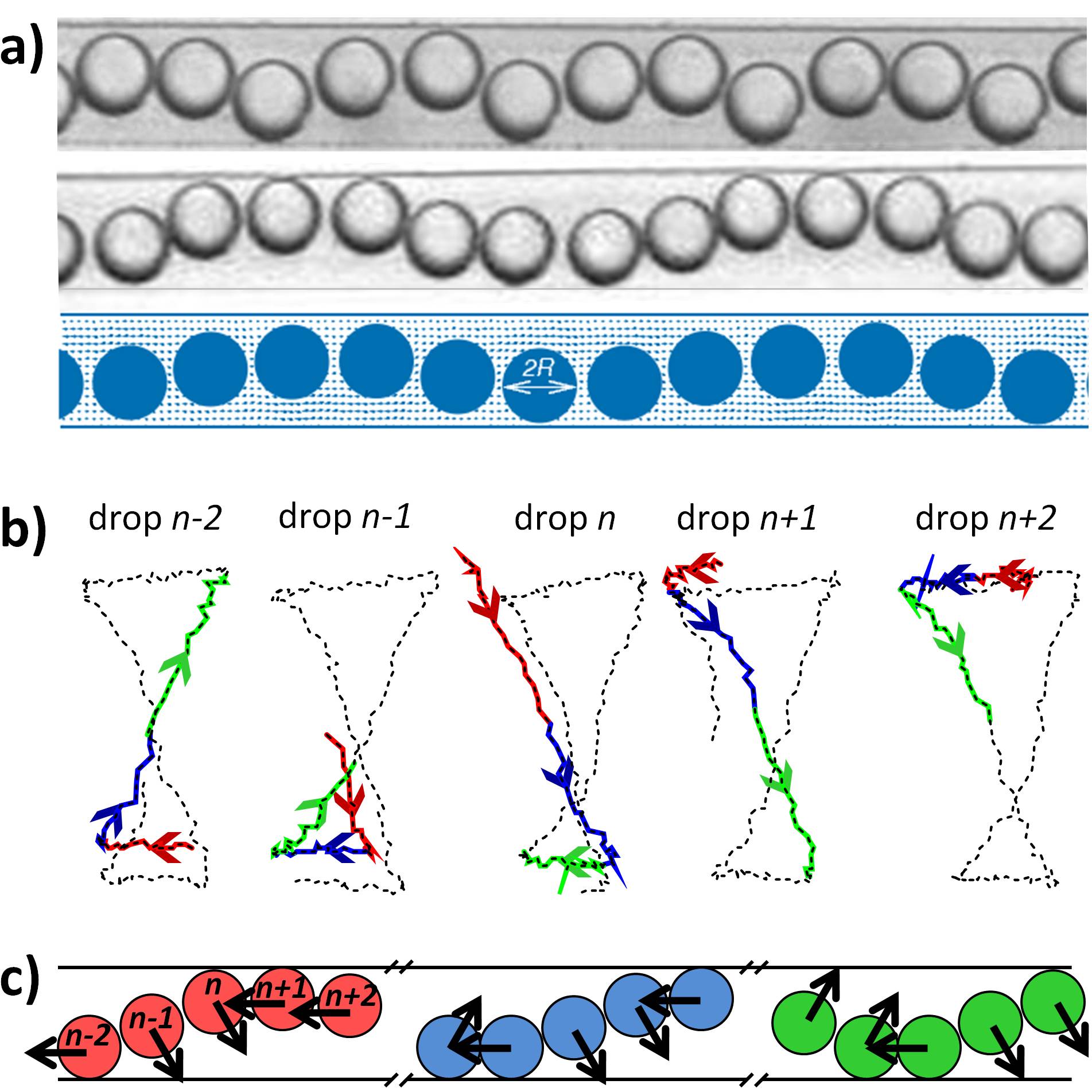}
  \caption{a) Traveling sine waves as generated by periodic defects
    for two different droplet sizes (\emph{top}) and by numerical
    simulation (\emph{bottom}) \cite{supplement}
    b) Configuration space trajectory of five neighboring droplets in the
    co-moving frame of the droplets as extracted from the experiments.  To
    avoid overlaps, the lateral distance between the trajectories was
    increased. The trajectories of three time intervals are colored in
    red (light gray), blue (dark gray), and green (gray).
    c) Sketch of the five droplets from b) using the same color code with
    indicated directions of motion in the co-moving frame of the droplets.}
  \label{fig:Picture2}
\end{figure}

Provided the droplet size relative to the channel width is chosen
appropriately, defects can be generated periodically. This leads to
sine-waves of droplets traveling forward in flow direction, see
Fig.~\ref{fig:Picture2}a. The collective oscillations are very stable
and could be observed for channel lengths up to 10\,cm, i.e. four
orders of magnitude larger than a typical droplet radius. The
wavelength $\lambda$ in longitudinal direction depends on both the
droplet size and the droplet spacing, i.e., droplet density. Varying
these parameters, various wavelengths can be specifically
excited. However, the experimentally accessible wavelength variation
is limited to a factor of about three as the size of the droplets and
the droplet density have to be chosen such that the droplets are close
enough to affect each other by hydrodynamic interactions but do not
deform each other by steric contact.

In the following, we analyze and discuss the specific experimental and
simulation results for a wavelength of six droplets with radius of
$R=64~\mu$m and crystal spacing $a=140~\mu$m. The results are equally valid
for other wavelengths and the discussion applies to both experiments and
simulation unless mentioned otherwise.

\emph{Analysis of Collective Oscillations}
The motion of five neighboring droplets is plotted in
Fig.~\ref{fig:Picture2}b,c in the co-moving frame of the droplets.
The space trajectory of each droplet describes a figure-eight and each
droplet has a constant phase shift to its neighboring droplets. The
figure-eight results from the transverse oscillation with comparable
large amplitude $W - 2R \approx 82~\mu m$ given by the lateral
confinement, and a coupled longitudinal oscillation with smaller
amplitude $a - 2R \approx 12~\mu m$ given by the crystal lattice,
i.e. the droplet density. For large transverse displacement, the
droplets closely approach the lateral channel walls, where they slow
down as a consequence of the no-slip boundary condition. Thus the
droplets move backward in the co-moving frame when they are close to a
channel wall, and forward when they cross the channel. One full cycle
of the longitudinal oscillation is completed while the droplet moves
from one wall to the other, corresponding to half a transverse
cycle. Accordingly, the constant phase shift between neighboring
droplets differs by a factor of two between transverse and for
longitudinal modes, i.e.  $55^{\circ}$ and $110^{\circ}$ for the
considered wave.

\begin{figure}
 \includegraphics[width=\columnwidth]{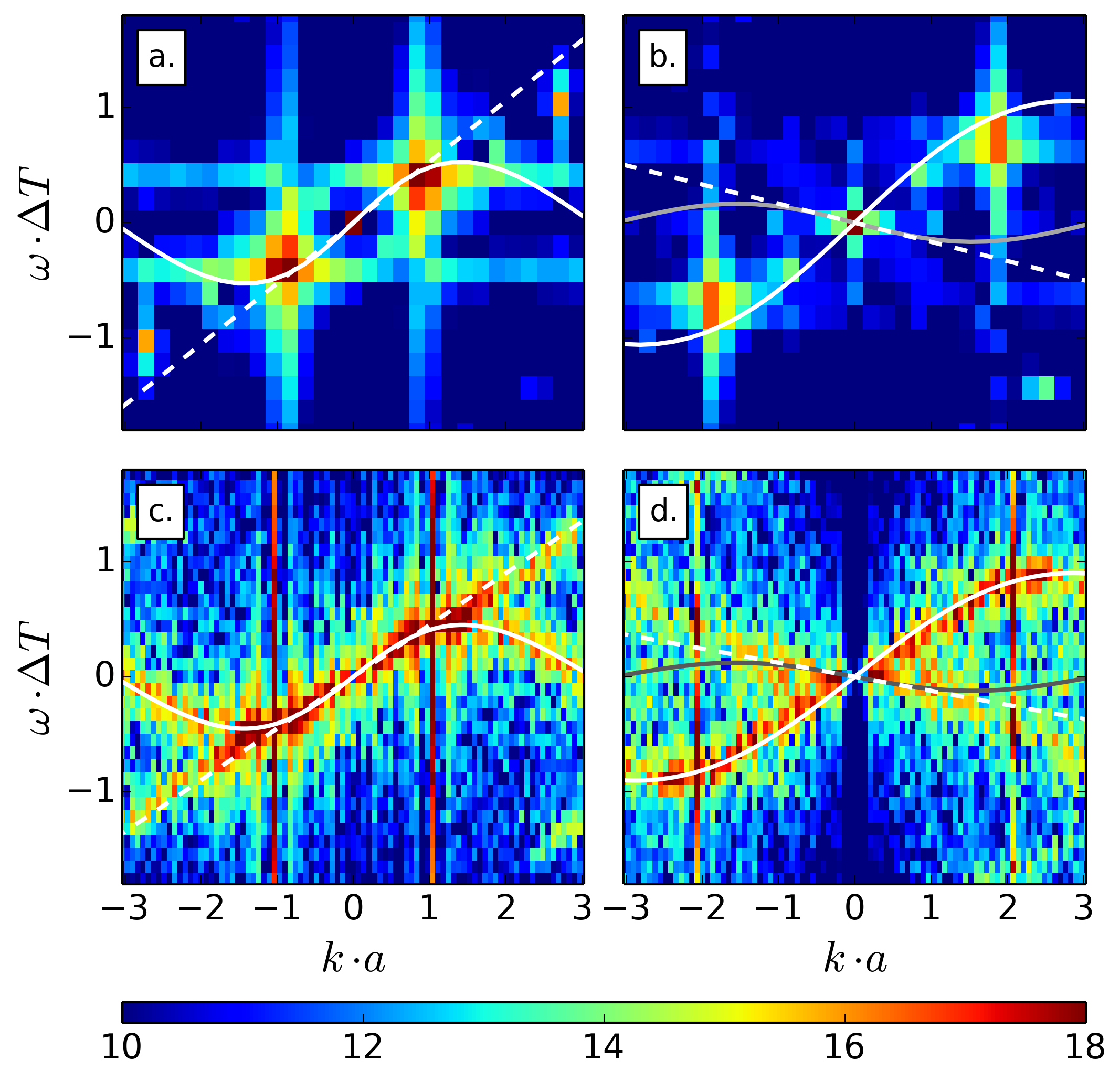}
  \caption{Power spectra of the Fourier transform of the droplet
    oscillations as extracted from experimental a,b) and simulation
    results c,d).
    a,c) Transverse modes $y(k, \omega)$; the white solid curves
    denote the prediction for acoustic phonons from the linearized theory
    Eq.~(\ref{eq:dispersion}) for $\omega_{\perp}(k)$ whereas the
    white dashed lines correspond to the continuum approximation with
    nearest neighbor interactions only.
    b,d) Longitudinal modes $x(k, \omega)$. The black/gray solid curves
    are the prediction $\omega_\parallel (k)$ from Eq.~(\ref{eq:dispersion})
    whereas the dashed lines correspond to the continuum approximation.
    The white solid curves represent $2 \omega_\perp(k/2)$ and illustrate
    the coupling of the longitudinal to the transverse modes. The time
    scale is $\Delta T=R/u_d$.}
  \label{fig:spectra}
\end{figure}

To analyze the droplet oscillations quantitatively, we extracted the
power spectra for the transverse and longitudinal oscillations both
from experimental and simulation results, cf. Fig.~\ref{fig:spectra}.
The main feature of the experimental results are a few distinct peak
values for both modes, cf. Fig.~\ref{fig:spectra}a,b.  The discrete
spectra demonstrate that the droplets describe a collective
oscillation with one well defined wavelength in both directions. The
corresponding phonon spectra obtained from simulations are displayed
in Fig.~\ref{fig:spectra}c,d. They show a continuous signature
revealing the full dispersion relation with a sine-like dependence of
the frequency on the wave vector. This complete signature results from
the small Peclet number, i.e., from the pronounced influence of
thermal fluctuations.

A comparison of the experimental and numerical transverse power
spectra shows that the dominant frequencies are identical. Moreover,
they are in excellent quantitative agreement with the dispersion
relation for transverse phonons $\omega_\perp(k)$ predicted by a
linearized far-field theory for two-dimensional channel
flow~\cite{Beatus2007,Beatus2012}:
\begin{equation}\label{eq:dispersion}
\begin{split}
\omega_{\perp}(k) &= 2 B \sum_{j=1}^{\infty} \sin(j k a) \left[ 1 + \cosh(j
\pi \beta) \right]^2 \mathrm{csch}^3(j \pi \beta)\\
\omega_{\parallel}(k) &= -4 B \sum_{j=1}^{\infty} \sin(j k a) \coth(j
\pi \beta) \mathrm{csch}^2(j \pi \beta),
\end{split}
\end{equation}
where $B=u_d/u_\text{oil}(u_\text{oil}-u_d)(\pi^2R/W^2)\tan(\pi R/W)$,
and $\beta=a/W$. Note that $\omega_{\perp}(k)$ in
Eq.~(\ref{eq:dispersion}) has a slightly different form than reported
in Refs.~\cite{Beatus2007, Beatus2012}, as it takes into account all
linear-order terms in the expansion \cite{supplement}.
The agreement confirms that the observed transverse oscillations are
acoustic phonons. From the dispersion relation, we obtain the
propagation speed $v_p\approx 312\,\mu\text{m/s}$ of the phonon mode
along the droplet crystal in direction of the flow.
The quantitative agreement also shows that the linearized far-field
hydrodynamic interactions describe the transverse phonon dynamics
remarkably well, despite the fact that the droplets nearly touch each
other \cite{Beatus2012}.
%
%\dcom{[RS] JB is hesitant about the former sentence as the considered
%situation / geometry is different to ours. Thus I propose an alternative
%sentence. If you are confident with the old sentence just change back to
%the old version.}
%\dreplace{Although it has been suggested that a stokeslet approximation is
%reasonable down to fairly small distances \cite{Cui2002}, it is not at
%all obvious that this still holds for dense droplet ensembles that are
%almost in contact when flowing in microchannels~\cite{Beatus2012}.}
%\dcancel{This is not at all obvious for the considered dense droplet
%ensembles that are almost in contact when flowing in
%microchannels~\cite{Beatus2012}.}
It even seems that the validity of the far-field approximation for an
ensemble of flat droplets is extended compared to an ensemble of
micro-particles with similar
confinement~\cite{Cui2002,Cui2004,Baron2008}.

\begin{figure}
  \includegraphics[width=\columnwidth]{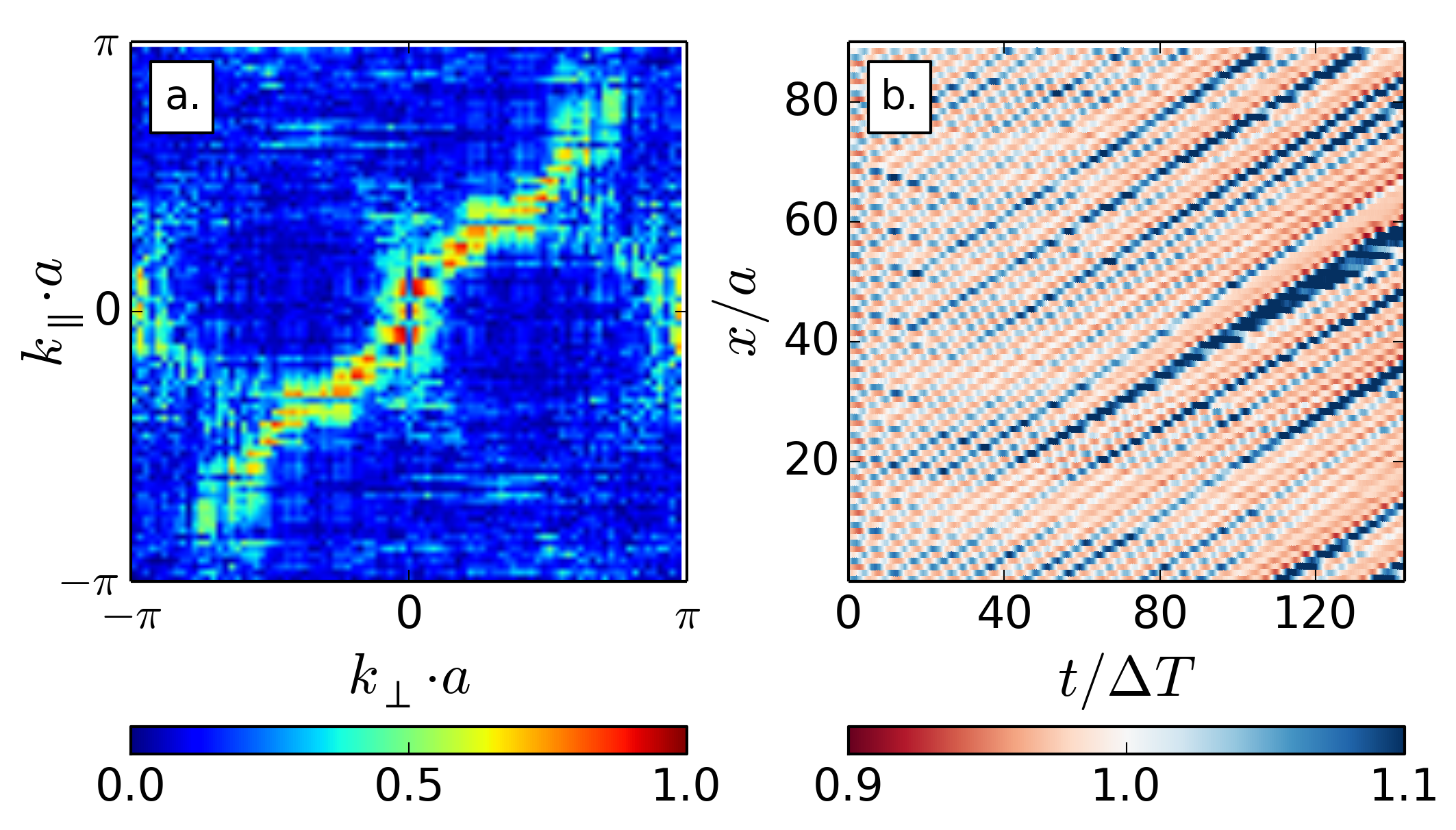}
  \caption{a) Correlation strength $C(k_\parallel,k_\perp)$ of
    longitudinal and transverse phonons as determined from the
    numerical simulation and b) Space-time diagram of the longitudinal
    droplet distance observed in simulation results.}
  \label{fig:correlations}
\end{figure}

The observed longitudinal modes are not actively excited and have a
smaller amplitude than the transverse oscillation. The dominant
frequencies derived from experimental and simulation result agree
quantitatively.  Interestingly, the longitudinal modes show a
dispersion relation that qualitatively deviates from the prediction of
the linearized theory, cf. Fig.~\ref{fig:spectra}b,d where
$\omega_\parallel(k)$ from Eq.~\ref{eq:dispersion} is plotted as gray
lines. Hence a different and non-linear mechanism must be responsible
for the longitudinal modes. We observe that the maximum of the
longitudinal dispersion relation appears shifted towards $k=\pi/a$,
and for all $k$, the longitudinal modes propagate in the flow
direction, in contrast to the analytical prediction
(\ref{eq:dispersion}). The observed shape of the longitudinal
dispersion relation resembles the scaled shape of the transverse
dispersion relation.
The apparent correlation is confirmed by the correlation strength
$C(k_\parallel,k_\perp) = {
\left\langle \tilde{x}(k_\parallel,t) \tilde{y}(k_\perp,t) \right\rangle_t } / { \sqrt{
\left\langle |\tilde{x}(k_\parallel,t)|^2 \right\rangle_t \left\langle
|\tilde{y}(k_\perp,t)|^2 \right\rangle_t } }$
between the numerically derived longitudinal and transverse modes
$\tilde{x}(k_\parallel,t)$ and $\tilde{y}(k_\perp,t)$, as displayed in
Fig.~\ref{fig:correlations}a. Whereas acoustic phonons are expected to
have high correlation for $k_\parallel=-k_\perp$ \cite{Liu2012}, we
observe high correlation for $k_\parallel\approx+k_\perp$.
In combination with the observed figure-eight motion of individual
droplets (Fig.~\ref{fig:Picture2}) it is suggested that the coupling
of the longitudinal to the transverse mode can be expressed by the
dispersion relation $\omega_{\parallel}(k)=2\omega_{\perp}(k/2)$. This
expression is plotted in Fig.~\ref{fig:spectra}b,d in quantitative
agreement with the observed longitudinal spectra.

\emph{Long-time behavior and onset of instability.}
Besides the main branches, a weaker and apparently linear branch is
present in the continuous numerical power spectra and consistent with
the secondary peaks in the experimental power spectra
(Fig.~\ref{fig:spectra}a,c). This branch has a positive slope for
transverse modes and a negative slope for longitudinal modes. In the
simulation results, the linear branches are relatively weak for short
simulation times but become more pronounced for longer times, which
indicates a relation to the long-time behavior of the droplets. The
space-time diagram of the numerically obtained longitudinal
elongations in Fig.~\ref{fig:correlations}b reveals the formation of
gaps and break-up of the crystal into smaller sub-units with
time. This is most likely caused by non-linear interactions between
longitudinal and transverse modes, similar to observations for
unconfined crystals \cite{Beatus2012} and in dusty-plasma crystals
\cite{Couedel2010,Liu2010}.  When the droplets move away from their
regular crystal positions, they can be regarded as a continuous
ensemble, and the finite-differences in the equation of motion are
replaced by a continuum approximation \cite{supplement}.  The
resulting linear dispersion relations are plotted as dashed lines in
Fig.~\ref{fig:spectra} and match the secondary features reasonably
well. They are more pronounced in the simulations as the breakup of
the crystal is amplified by the stronger thermal fluctuations. Because
of the longitudinal modes, fluctuations in the longitudinal droplet
positions are also present in the experiments. However, as the
amplitude of the longitudinal fluctuations are not growing on the
observed length and time scales, the peaks remain faint in the
experimental power spectra. This experimental observation seems to
indicate that fluctuations in a dense microfluidic crystal with small
droplet spacing are more stable than a dilute microfluidic crystal
with large droplet spacing~\cite{Desreumaux2012}.

\emph{Summary and Conclusion.}
Coupled phonon modes were experimentally studied in a dense
one-dimensional microfluidic crystal and quantitatively compared to
simulations and analytical results. The specifically excited
transverse oscillations can be described as acoustic phonons by a
linearized far-field theory even in the dense system where droplets
are almost in contact.
The considerable amplitude of the excited transverse oscillations
leads to a non-linear coupling of longitudinal to transverse modes.
The coupling is due to the lateral variation of the hydrodynamic
interactions across the channel. The coupled longitudinal modes are
beyond the existing analytic description, but can be quantitatively
described by a phenomenological dispersion relation.
The small longitudinal droplet spacing in dense droplet crystals
seems to support the coupling of the longitudinal and the transverse
modes leading to very stable collective oscillations.
%
%Thus dense microfluidic crystals are a promising system to study
%collective effects and non-linear phenomena arising from hydrodynamic
%interactions under confinement which could not yet be observed in
%dilute systems.
%
Thus, complementary to previous studies on dilute systems, dense
microfluidic crystals are a promising system to study collective
effects and non-linear phenomena arising from hydrodynamic
interactions under confinement. In particular, it would be interesting
to extend this study to two-dimensional systems, where higher order
collective excitations and unexplored self-organizing behavior may
arise.

%\dcancel{The effects we observe may have general impact on dynamic
%pattern formation in confined microflows
%\mbox{\cite{Thorsen2001,McWhirter2009}} and have potential
%applications in the design of chips to control droplet motion, e.g.,
%in analytic essays~\mbox{\cite{Huebner2008,Koester2008,Schmitz2009}}.
%%
%Using this microfluidic system also athermal transitions in a driven
%non-equilibrium crystal or XXX could be studied.}

R.S. and J.-B. F. gratefully acknowledge financial support by the DFG
grant SE 1118/4.

%%%%%%%%%%%%%%%%%%%%%%%%%%%%%%%%%%%%%%%%%%%%%%%%%%%%%%%%%%%%%%%%%%%%%%%%%%%%%%

%\bibliography{droplets}

%\bibitem{supplement} See supporting information at [URL will be inserted by publisher].


%%%%%%%%%%%%%%%%%%%%%%%%%%%%%%%%%%%%%%%%%%%%%%%%%%%%%%%%%%%%%%%%%%%%%%%%%%%%%%

\end{document}